\documentclass{article}
\usepackage{graphicx} 
\usepackage{amsmath,amssymb,bm}
\usepackage{booktabs}
\usepackage[margin=2.5cm]{geometry}
\usepackage{xcolor}
\usepackage{hyperref}
\definecolor{citationnavy}{RGB}{20,45,78}
\hypersetup{
    colorlinks=true,
    citecolor=citationnavy,
    linkcolor=black,
    urlcolor=black
}
\numberwithin{equation}{section}

\newcommand{\rc}{r_{\rm c}}
\newcommand{\Mc}{M_{\rm c}}
\newcommand{\PGW}{P_{\rm GW}}
\newcommand{\Pint}{P_{\rm int}}
\newcommand{\Rt}{\widetilde R}
\newcommand{\dd}{\mathrm{d}}
\newcommand{\ket}[1]{|#1\rangle}
\newcommand{\safefig}[2]{\IfFileExists{#2}{\includegraphics[width=#1]{#2}}{\fbox{\parbox[c][4cm][c]{#1}{\centering\texttt{\detokenize{#2}}\\ (figure file not uploaded)}}}}

\title{Ionization of gravitational atoms\\ by a companion-sourced Yukawa potential}

\author{%
\makebox[\textwidth][c]{%
\begin{minipage}{0.95\textwidth}
\centering
Yi Wang$^{b,c}$ \quad Jiayi Yao$^{a,b,c}$\\[0.8em]
{\small $^{a}$ School of Space Science and Technology, Shandong University, Weihai 264209, China}\\[0.3em]
{\small $^{b}$ Department of Physics, The Hong Kong University of Science and Technology, Clear Water Bay, Kowloon, Hong Kong S.A.R., P.R. China}\\[0.3em]
{\small $^{c}$ Jockey Club Institute for Advanced Study, The Hong Kong University of Science and Technology, Hong Kong S.A.R., P.R. China}\\[0.3em]
{\small E-mail: \texttt{phyw@ust.hk}, \texttt{202300202037X@mail.sdu.edu.cn}}
\end{minipage}}%
}

\date{}

\begin{document}

\maketitle

\begin{abstract}
Superradiant instabilities around rotating black holes can produce clouds of ultralight bosons bound by the black hole's gravitational field. A binary companion can ionize these clouds by driving transitions into continuum states in the black hole's $1/r$ gravitational potential, extracting energy and angular momentum from the orbit. We investigate how a companion-sourced Yukawa interaction modifies this process. At weak coupling, the interference term between the gravitational and Yukawa transition amplitudes provides the leading correction to the ionization power. For the orbital frequencies here, an attractive interaction enhances continuum ionization and accelerates the inspiral,
whereas a repulsive interaction has the opposite effect. At larger coupling, the pure Yukawa contribution to the ionization power becomes increasingly important and can compete with the gravitational contribution, particularly for lighter mediators. Comparing continuum ionization with off-resonant bound-state mixing, we find that continuum ionization can halt superradiant growth at a substantially smaller Yukawa coupling. At their respective critical couplings, the Yukawa interaction provides the dominant contribution to cloud depletion in both channels. Finally, we calculate the phase correction to gravitational waves due to continuum ionization and characterize how the chirp-rate response varies with frequency near the thresholds at which successive orbital harmonics open continuum-ionization channels.
\end{abstract}

\tableofcontents

\section{Introduction}
\label{sec:intro}
A rotating Kerr black hole can transfer its rotational energy and angular momentum to bosonic waves through superradiance. This amplification takes place in the ergoregion. A bosonic mode that co-rotates with the black hole is amplified when the superradiant condition is satisfied, extracting rotational energy and angular momentum from the black hole. For a massive boson, the mass term provides an effective barrier at distances of order its Compton wavelength. The amplified wave cannot propagate freely to infinity and is effectively reflected back toward the black
hole, where it can undergo superradiant amplification again. Through this repeated amplification, the bosonic mode grows exponentially and a large number of bosons gradually accumulate around the black hole, forming a macroscopic cloud ~\cite{Arvanitaki2011,Brito2015,Detweiler1980,TWZ2022}. The cloud is bound by the $1/r$ gravitational potential of the black hole, so its non-relativistic spectrum is that of hydrogen, and the black hole together with its cloud is often called a gravitational atom. 

A binary companion perturbs the atom. Its tidal field can mix bound levels resonantly~\cite{Baumann2019,Baumann2020,Berti2019}, and it can also drive transitions from bound states into the continuum of the same $1/r$ potential. This ionization has been worked out by Baumann, Bertone, Stout and Tomaselli (BBST)~\cite{BBST2022,BBST2022prl}, who showed that the energy it removes from the orbit can exceed the gravitational-wave luminosity by about two orders of magnitude. Tomaselli, Spieksma and Bertone~\cite{TSB2023} identified the orbital backreaction of ionization as dynamical friction~\cite{Traykova2021}, corrected the treatment of the gravitational dipole, and extended the calculation to eccentric and inclined orbits. Related applications of bound-to-continuum transitions to stellar environments around supermassive black holes have recently been studied in Ref.~\cite{Cheng2026}. Tong, Wang and Zhu~\cite{TWZ2022} showed that a companion can suppress superradiant growth by mixing a superradiant state with a strongly absorptive state. The continuum states considered here are those of the black hole's $1/r$ gravitational potential, with the Yukawa potential treated as a perturbation.

Beyond the standard gravitational interaction, additional forces can be mediated by massive fields, giving rise to modifications of the gravitational potential. Thus, we consider an additional Yukawa interaction between the companion and the cloud particles, alongside Newtonian gravity. The total potential is
\begin{equation}
  V(t,\bm r)=-\frac{q\alpha}{|\bm r-\bm R(t)|}+V_X(t,\bm r),
\end{equation}
the Yukawa potential is
\begin{equation}
  V_X(t,\bm r)=-\beta q\alpha\,\frac{e^{-m_X|\bm r-\bm R(t)|}}{|\bm r-\bm R(t)|},
  \label{eq:VX}
\end{equation}
where $q=M_*/M$ is the mass ratio and $\bm R(t)$ the position of the companion. This is the parametrization used in fifth-force searches~\cite{Adelberger2003}: $\beta$ is the strength of the Yukawa force relative to gravity at distances short compared with its range $\lambda=m_X^{-1}$. A spin-independent Yukawa force could arise from a light scalar coupled to the companion and the cloud particles, or from CP-violating axion couplings~\cite{OHare2020}. Scalar-mediated Yukawa interactions between ultralight dark matter particles have been studied in
the context of black hole superradiance~\cite{Lyu2025}.
Such an interaction could give rise to a companion-sourced
Yukawa potential if the companion contains dark matter
coupled to the same mediator. In contrast, single-axion exchange in a minimal CP-conserving QCD-axion model is spin dependent. Background-enhanced two-axion forces are a separate mechanism~\cite{Grossman2025}. We make four assumptions: (i) $\beta$ is a signed parameter, with $\beta>0$ attractive and $\beta<0$ repulsive; (ii) the black hole carries no Yukawa charge, so the Yukawa dipole is not cancelled by its acceleration (section~\ref{sec:multipole}); (iii) the companion is treated as a point source with a spin-independent interaction; (iv) $\beta$ describes the interaction between the companion and the cloud particle. Laboratory fifth-force bounds cannot be applied directly to this $\beta$ without specifying how the relevant couplings are related\footnote{Laboratory fifth-force bounds cannot be applied directly to $\beta$ for two reasons. First, the Yukawa coupling of the bosons that form the cloud is unknown and is not measured by fifth-force experiments using ordinary matter. For example, a scalar-mediated
fifth force may act exclusively within the dark
sector~\cite{Lyu2025}. Second, if the companion is a neutron star, finite-density effects can give it an effective charge per unit mass different from that of terrestrial test bodies~\cite{HookHuang2018}.}.

The remainder of this paper is organized as follows: Section 2 introduces the hydrogenic variables, continuum normalization, and the expressions for transition rates, orbital power and torque. Section 3 develops the multipole expansion of the companion-sourced gravitational and Yukawa potentials in the black-hole frame, and calculates the interference between their ionization amplitudes. Section 4 presents the results at fixed orbital separation, including the dependence on mediator mass and the sign of $\beta$, as well as cloud depletion. Section 5
compares continuum ionization with the off-resonant $211\to200$ transition, first considering the Yukawa interaction alone and then including gravitational contributions and interference. Section 6 uses the ionization power to determine the orbital evolution
and the associated phase correction to gravitational waves. Section 7 summarizes our conclusions. Appendix A gives the angular and radial matrix elements, Appendix B describes the numerical method and its validation, Appendix C estimates the effects of the static $g=0$ Yukawa harmonic, and Appendix D discusses the domain of validity.

\section{Conventions \& Review of the continuum states}
\label{sec:conventions}
This section reviews the non-relativistic bound and continuum states of the black hole's $1/r$ gravitational potential, together with the standard treatment of bound-to-continuum transitions. We adopt the normalization of continuum states used by BBST~\cite{BBST2022} and follow their treatment of the ionization rate. Ref.~\cite{Cheng2026} provides a related application to stellar perturbations around supermassive black holes.

\subsection{Hydrogenic variables}
We introduce the variables used to describe the hydrogenic cloud and the companion's circular orbit, and specify the non-relativistic bound state considered in this work.

We use $G=c=\hbar=1$. The black-hole mass is $M$, the boson mass $\mu$, and
\begin{equation}
  \alpha=M\mu,\qquad \rc=\frac{1}{\mu\alpha}=\frac{M}{\alpha^2},\qquad
  \rho=\frac{r}{\rc},\qquad x=\frac{R}{\rc}=\frac{\alpha^2R}{M}.
  \label{eq:alpha}
\end{equation}
Here, $\alpha$ is the gravitational fine-structure constant, and $r_{\rm c}$ is the Bohr radius of the gravitational atom. The variables $\rho$ and $x$ denote the dimensionless radial coordinate and orbital separation, respectively.

For a circular binary with mass ratio $q=M_*/M$, Kepler's law $\Omega^2=M(1+q)/R^3$ gives
\begin{equation}
  \omega(x)\equiv\frac{\Omega}{\mu\alpha^2}=\frac{\sqrt{1+q}}{x^{3/2}} .
  \label{eq:omega}
\end{equation}
We define $q$ as the mass ratio of the companion to the central black hole, and $\Omega$ as the orbital angular frequency. The corresponding dimensionless frequency is $\omega(x)$.

The unperturbed non-relativistic Hamiltonian is
\begin{equation}
  H_0=-\frac{\nabla^2}{2\mu}-\frac{\alpha}{r},
  \label{eq:H0}
\end{equation}
and an initial state $\ket b=\ket{n\ell m}$ has energy
\begin{equation}
  \epsilon_b=-\frac{\mu\alpha^2}{2n^2}
  \label{eq:eps}
\end{equation}
at leading order. We work with $\ket{211}$. Dimensionless radial functions are defined by $R_{n\ell}(r)=\rc^{-3/2}\Rt_{n\ell}(\rho)$, so that
$\Rt_{21}(\rho)=\rho\,e^{-\rho/2}/(2\sqrt6)$ and $\langle\rho\rangle_{21}=5$.

\subsection{Continuum normalization}
We introduce the continuum states that serve as final states in the ionization process, specifying their normalization and radial wavefunctions. Energy conservation determines the momentum of each final state and the conditions under which the corresponding
ionization channel opens. We also discuss the behavior of the continuum wavefunctions near the channel threshold.

The continuum states satisfy
\begin{equation}
  \langle k;\ell m|k';\ell'm'\rangle=2\pi\,\delta(k-k')\,\delta_{\ell\ell'}\delta_{mm'} .
  \label{eq:norm}
\end{equation}
Their radial functions are
\begin{equation}
  R_{k\ell}(r)=\frac{2}{r}\,F_\ell\!\left(-\frac{\mu\alpha}{k},\,kr\right),
  \label{eq:Rkl}
\end{equation}
where $F_\ell(\eta_{\rm C},z)$ is the regular radial solution associated with the attractive $1/r$ potential, with $\eta_{\rm C}=-\mu\alpha/k$. In dimensionless form $R_{k\ell}(r)=\rc^{-1}\Rt_{\kappa\ell}(\rho)$ with
\begin{equation}
  \kappa=k\rc=\frac{k}{\mu\alpha},\qquad
  \Rt_{\kappa\ell}(\rho)=\frac{2}{\rho}F_\ell\!\left(-\frac1\kappa,\kappa\rho\right).
  \label{eq:kappa}
\end{equation}
Energy conservation, $\epsilon_b+g\Omega=k_*^2/2\mu$, fixes the momentum of the final state reached through orbital harmonic $g$:
\begin{equation}
  \kappa_g^2=2\left[g\,\omega(x)-\frac{1}{2n^2}\right].
  \label{eq:kappag}
\end{equation}
The channel is open only when $\kappa_g^2>0$. For $n=2$ this means $g\,\omega(x)>1/8$, i.e.\ $R<R_*^{(g)}$; as the companion spirals in, the high harmonics open first and $g=1$ opens last.

Near threshold, the continuum radial wavefunction scales as $k^{1/2}$:
\begin{equation}
  R_{k\ell}(r)\xrightarrow{k\to0}
  \sqrt{\frac{4\pi k}{r}}\,
  J_{2\ell+1}\left(2\sqrt{2\mu\alpha r}\right).
  \label{eq:zero}
\end{equation}
A free wave with the same normalization instead vanishes as $k^{\ell+1}$. The asymptotic form in eq.~\eqref{eq:zero} follows from the Kummer representation of the continuum radial solution; details are given in appendix~\ref{app:matrix}.
Consequently, the bound-to-continuum matrix elements considered here scale as $\sqrt{k}$, so the combination $|\eta|^2/k$ entering the rate remains finite at threshold. Each harmonic therefore produces a finite step in the ionization rate when
it opens at fixed $\Omega$. This step is a feature of the golden-rule approximation at fixed orbital frequency. During an inspiral, $g\Omega$ sweeps through the threshold, and the finite energy resolution of the transition, $\Delta(g\Omega)\sim\sqrt{\partial_t(g\Omega)}$, smooths the step over this interval
(section~\ref{sec:obs-degeneracy}). The finite local spectral weight at threshold is a manifestation of Sommerfeld enhancement due to the attractive $1/r$ gravitational potential and is consistent with the accumulation of high-$n$ hydrogenic bound states near the continuum threshold.

\subsection{Rates, power and torque}
A perturbation $V$ with harmonic components $V^{(g)}e^{-ig\Omega t}$ couples $\ket b$ to the continuum state $\ket{K_*}=\ket{k_*;\ell_fm_f}$ with matrix element $\eta^{(g)}_{K_*b}=\langle K_*|V^{(g)}|b\rangle$. Fermi's golden rule with the measure $\int\dd k/2\pi$ implied by eq.~\eqref{eq:norm} gives the occupation-number loss rate~\cite{BBST2022}
\begin{equation}
  \Gamma_N\equiv-\frac{\dd\ln|c_b|^2}{\dd t}
  =\sum_{\ell_f,g}\frac{\mu\,\bigl|\eta^{(g)}_{K_*b}\bigr|^2}{k_*^{(g)}}\;\Theta\!\left[\bigl(k_*^{(g)}\bigr)^2\right].
  \label{eq:rate}
\end{equation}
This is an occupation-number rate; the amplitude $c_b$ decays at half of it. Each ionized particle takes energy $g\Omega$ from the orbit, so with $N=\Mc/\mu$ particles in the cloud the ionization power is
\begin{equation}
  P_{\rm ion}=\frac{\Mc}{\mu}\sum_{\ell_f,g}g\Omega\,
  \frac{\mu\,\bigl|\eta^{(g)}_{K_*b}\bigr|^2}{k_*^{(g)}}\;\Theta\!\left[\bigl(k_*^{(g)}\bigr)^2\right].
  \label{eq:power}
\end{equation}
For a co-rotating orbit, each ionized quantum carries away energy $g\Omega$ and azimuthal angular momentum $g$ (in units of $\hbar$) from the orbit. Since a circular orbit satisfies $dE_{\rm orb}=\Omega\,dL_{\rm orb}$, the orbit evolves along the circular sequence, with
\begin{equation}
\dot E_{\rm orb}=-P_{\rm ion},
\qquad
\dot L_{\rm orb}=-\frac{P_{\rm ion}}{\Omega}.
  \label{eq:orbital_drag}
\end{equation}
Equation~\eqref{eq:orbital_drag} describes the dissipative drag due to continuum ionization, distinct from the floating or sinking torque generated by coherent bound--bound absorption \cite{TWZ2022}. We express all powers in units of the quadrupole luminosity of a circular binary,
\begin{equation}
P_{\rm GW}
=
\frac{32}{5}q^2(1+q)
\left(\frac{M}{R}\right)^5
=
\frac{32}{5}q^2(1+q)\frac{\alpha^{10}}{x^5}.
  \label{eq:PGW}
\end{equation} 
Here, $M_c=M_c(t)$ denotes the cloud mass during the inspiral. The rate in eq.~(2.10) is obtained in the Markov (Weisskopf--Wigner) approximation. For the benchmark used in section~6, the orbital evolution is adiabatic. Including gravitational ionization,
we find $\dot{\Omega}/\Omega^2\simeq 10^{-5}$ at $R=150M$. For the dominant harmonics, the occupation-number loss rate $\Gamma_N$ is also smaller than the chirp-induced energy width $\Delta(g\Omega)$. We neglect accretion of cloud particles onto the companion. The expressions in eqs.~(2.10)--(2.12) apply to a general perturbation. Below, we use them for the gravitational and Yukawa interactions, with their amplitudes added coherently.

\section{Companion-sourced Yukawa ionization}
\label{sec:yukawa}
We construct the gravitational and Yukawa perturbations sourced by the companion in the black-hole frame and derive the corresponding ionization amplitudes and their interference.

\subsection{Multipole expansion and the black-hole frame}
\label{sec:multipole}
The $L\geq 2$ multipoles of the Newtonian potential generated by a companion on an equatorial orbit are given by~\cite{BBST2022}:
\begin{equation}
  V_g(t,\bm r)=-\frac{q\alpha}{\rc}\sum_{L\ge2,M}\frac{4\pi}{2L+1}\,
  Y^*_{LM}\!\left(\frac\pi2,\varphi_*(t)\right)Y_{LM}(\hat{\bm r})\,
  \frac{\rho_<^L}{\rho_>^{L+1}} .
  \label{eq:Vg}
\end{equation}
Here, $\rho_< = \min(\rho,x)$, $\rho_> = \max(\rho,x)$, and $\gamma$ is the angle
between $\hat{\boldsymbol r}$ and $\hat{\boldsymbol R}$. The term with azimuthal number $M$ has time dependence $e^{-iM\phi_*(t)}$. For a co-rotating orbit, $\phi_*(t)=\Omega t$, so the component with $M=g>0$ drives the $g$-th harmonic. The angular integral
imposes the selection rule $m_f=m_b+g$. The $M=0$ components are static and cannot open an ionization channel. Indeed, setting $g=0$ in eq.~(2.8) gives $\kappa_0^2=-1/n^2<0$. The static components of the Yukawa potential can still shift and dress the bound
state. Their effects are estimated in appendix~C.

The dipole term needs care because the cloud is described in the frame of the black hole, which is accelerated by the companion. In that frame each cloud particle feels the inertial potential $+\mu\,\bm a_{\rm BH}\cdot\bm r=(q\alpha/\rc)(\rho/x^2)P_1(\cos\gamma)$. Adding it to the $L=1$ term of the Newtonian expansion gives the corrected dipole contribution used in ref.~\cite{TSB2023},
\begin{equation}
  K^g_1(\rho,x)=\frac{\rho_<}{\rho_>^2}-\frac{\rho}{x^2}
  =\left(\frac{x}{\rho^2}-\frac{\rho}{x^2}\right)\Theta(\rho-x).
  \label{eq:dipole}
\end{equation}
Inside the orbit, the dipole field is uniform and is canceled by the acceleration of the black hole. Outside the orbit, it survives. We use eq.~(3.2) for $L=1$ and eq.~(3.1) for $L\geq2$. Since $L=1$, the dipole contributes only to harmonics with $|g|\leq1$.

With the range parameter
\begin{equation}
  \zeta=m_X\rc=\frac{m_X/\mu}{\alpha},
  \label{eq:zeta}
\end{equation}
the spatial dependence of the Yukawa potential can be expanded in multipoles as
\begin{equation}
  \frac{e^{-\zeta|\bm\rho-\bm x|}}{|\bm\rho-\bm x|}=\sum_{L=0}^\infty K^X_L(\rho,x;\zeta)\,P_L(\cos\gamma),
  \qquad
  K^X_L=(2L+1)\,\zeta\,i_L(\zeta\rho_<)\,\hat k_L(\zeta\rho_>),
  \label{eq:KX}
\end{equation}
with\footnote{Our normalization is related to the convention used by some numerical libraries by $\hat{k}_L(z)=(2/\pi)k_L^{\mathrm{lib}}(z)$, where $k_L^{\mathrm{lib}}$ denotes the modified spherical Bessel function of the second kind in that convention.} 
\begin{equation}
  i_L(z)=\sqrt{\frac{\pi}{2z}}\,I_{L+1/2}(z),\qquad
  \hat k_L(z)=\sqrt{\frac{2}{\pi z}}\,K_{L+1/2}(z),\qquad \hat k_0(z)=\frac{e^{-z}}{z}.
  \label{eq:bessel}
\end{equation}

Equation~\eqref{eq:KX} follows from the radial Green function of $\nabla^2-m_X^2$. The function $i_L$ is regular at the origin, whereas $\hat{k}_L$ decays at infinity. Their product therefore gives the radial Green function, with the Wronskian
\begin{equation}
W[i_L,\hat{k}_L](z)=-\frac{1}{z^2}
  \label{eq:wronskian}
\end{equation}
fixing the normalization $(2L+1)\zeta$.
In the long-range limit, $\zeta\to0$, the radial multipole function reduces to 
\begin{equation}
K_L^X\longrightarrow
\frac{\rho_<^L}{\rho_>^{L+1}},
  \label{eq:yukawa_longrange}
\end{equation}
recovering the corresponding Newtonian multipole. In the opposite limit, $\zeta\gg1$, the Yukawa potential becomes short-ranged and approaches a contact interaction in the distributional sense. For the smooth bound--continuum overlap at fixed $x$, the resulting Yukawa transition amplitude scales as $\zeta^{-2}$.

The radial integrals are
\begin{equation}
  I^{g,X}_{\ell_fL}=\int_0^\infty\dd\rho\,\rho^2\,\Rt_{\kappa\ell_f}(\rho)\,\Rt_{n\ell_b}(\rho)\,K^{g,X}_L(\rho,x),
  \label{eq:IX}
\end{equation}
where $K_L^X$ is the radial function defined in eq.~(3.4). For the gravitational potential, the corresponding radial function is $K_L^g=\rho_<^L/\rho_>^{L+1}$ for $L\geq2$, while the $L=1$ contribution is given by eq.~(3.2). Combining the powers of $r_c$, from the volume element ($r_c^3$), continuum wavefunction ($r_c^{-1}$), bound-state wavefunction ($r_c^{-3/2}$) and potential ($r_c^{-1}$), the matrix elements are
\begin{equation}
  \eta^{(g)}_g=-q\alpha\,\rc^{-1/2}\mathcal A^{(g)}_g,\qquad
  \eta^{(g)}_X=-q\alpha\,\rc^{-1/2}\mathcal A^{(g)}_X,
  \label{eq:etaX}
\end{equation}
where $\mathcal A^{(g)}_g$ is given by eq.~\eqref{eq:A3} of appendix~\ref{app:matrix} and $\mathcal A^{(g)}_X$ by the same expression with $I^g_{\ell_fL}$ replaced by $I^X_{\ell_fL}$. Here $\eta_X$ is the Yukawa matrix element per unit $\beta$.

In the limit $\zeta\to0$, the Yukawa radial multipole functions approach the corresponding Newtonian radial functions, including the uncorrected $L=1$ term, whereas $V_g$ uses the corrected dipole. The two therefore agree only where $g=1$ is closed. This is a useful numerical consistency check: at $x=6$ and $\zeta=0.01$ we find $P_X/P_g=0.9990$. At $x=2.454$, where $g=1$ is open, the unsubtracted dipole gives a rate $6.1$ times larger than the corrected one.

\subsection{Gravity--Yukawa interference}
\label{sec:interference}
For a fixed final state the amplitudes add before squaring,
\begin{equation}
  \eta^{(g)}_{\rm tot}=\eta^{(g)}_g+\beta\,\eta^{(g)}_X .
  \label{eq:etatot}
\end{equation}
Inserting this into eqs.~\eqref{eq:rate} and~\eqref{eq:power} gives
\begin{equation}
  P_{\rm ion}=P_g+\beta\,\Pint^{(1)}+\beta^2P_X^{(1)},
  \label{eq:Ptot}
\end{equation}
with
\begin{equation}
  \bigl\{P_g,\ \Pint^{(1)},\ P_X^{(1)}\bigr\}=\frac{\Mc}{M}\,q^2\alpha^5\sum_{\ell_f,g}\frac{g\,\omega(x)}{\kappa_g}
  \Bigl\{|\mathcal A_g|^2,\ 2\,{\rm Re}\bigl(\mathcal A_g\mathcal A_X^*\bigr),\ |\mathcal A_X|^2\Bigr\},
  \label{eq:Pparts}
\end{equation}
and the same decomposition $\Gamma_N=\Gamma_{N,g}+\beta\Gamma_{N,{\rm int}}+\beta^2\Gamma_{N,X}$ for the rate, without the weight $g\omega$. The superscript $(1)$ means the coefficient is evaluated at $\beta=1$; we drop it where no confusion arises. Two derived quantities summarize the comparison with gravity:
\begin{equation}
  \beta_{\rm eq}^{P}\equiv
  \left(\frac{P_g}{P_X}\right)^{1/2},
  \qquad
  c_P\equiv
  \frac{\Pint}{2\sqrt{P_gP_X}},
  \qquad
  \frac{P_{\rm ion}}{P_g}
  =
  1+2c_P\frac{\beta}{\beta_{\rm eq}^{P}}
  +\left(\frac{\beta}{\beta_{\rm eq}^{P}}\right)^2 .
  \label{eq:ceq}
\end{equation}
The coupling $\beta_{\rm eq}^{P}$ is the value at which the Yukawa and gravitational amplitude vectors have equal norms. The coefficient $c_P$ measures their normalized overlap over the open final states and satisfies $|c_P|\leq 1$ by the Cauchy--Schwarz inequality. For $c_P>0$ and $\beta<0$ the power falls below $P_g$ for $|\beta|<\Pint/P_X$ and reaches its minimum
\begin{equation}
  P_{\rm min}=P_g(1-c_P^2)
  \qquad\text{at}\qquad
  \beta_*=-\frac{\Pint}{2P_X}
  =-c_P\beta_{\rm eq}^{P}.
  \label{eq:Pmin}
\end{equation}
Treating $P_g$ and $P_X$ as independent positive rates would miss both the linear regime and this suppression.

\section{Yukawa results at fixed separation}
\label{sec:results}
Using these amplitudes, we evaluate the ionization power at fixed orbital separation and examine its dependence on the mediator mass and coupling sign, as well as its
implications for cloud depletion.

\subsection{Benchmark and power decomposition}
\label{sec:benchmark}
We adopt the BBST benchmark of ref.~\cite{BBST2022}, 
\begin{equation}
\alpha=0.2,\qquad q=10^{-3},\qquad
\frac{M_c}{M}=10^{-2},\qquad
|b\rangle=|211\rangle,
\label{eq:bench}
\end{equation}
for a co-rotating orbit. The first five ionization thresholds, obtained from eq.~\eqref{eq:kappag}, are
\begin{equation}
\frac{R_*^{(g)}}{M}
=
\alpha^{-2}\left[4g^2(1+q)n^4\right]^{1/3}
=
100.03,\ 158.79,\ 208.08,\ 252.07,\ 292.50,
\quad g=1,\ldots,5.
\label{eq:thresholds}
\end{equation}
Our numerical implementation reproduces the BBST gravitational ionization curve and these thresholds (appendix~\ref{app:reproduction}).
At the reference separation $R/M=150$, corresponding to $x=6$ and $r_c=25M$, the $g=1$ channel is closed, while all $g\geq2$ channels are open. The $g=2$ and $g=3$ channels account for approximately 70\% and 21\% of the gravitational ionization power, respectively. Since the dipole contributes only to $|g|\leq1$, it does not enter the continuum ionization rate at this separation. The gravitational baseline is therefore
\begin{equation}
\frac{P_g}{P_{\rm GW}}=147.277,
\label{eq:grav_benchmark}
\end{equation}
in agreement with BBST. We use this reference point to compare the power decomposition for three mediator masses in table~\ref{tab:decomp}.

\begin{table}[htbp]
\centering
\small
\begin{tabular}{ccccccccc}
\toprule
$m_X/\mu$ & $P_g$ & $P_X$ & $\Pint$ & $P_{\rm ion}(\beta{=}1)$ & $c_P$ & $\beta_{\rm eq}^{P}$ & $\Pint/P_X$ & $P_{\rm min}/P_g$\\
\midrule
1.00 & 147.277 & 0.0319 & 4.062 & 151.371 & 0.937 & 67.95 & 127.4 & 12.2\% \\
0.20 & 147.277 & 8.993 & 70.613 & 226.882 & 0.970 & 4.05 & 7.85 & 5.9\% \\
0.05 & 147.277 & 93.236 & 233.945 & 474.458 & 0.9982 & 1.26 & 2.51 & 0.36\% \\
\midrule
0.20 (counter) & 149.397 & 27.906 & 125.662 & 302.965 & 0.973 & 2.31 & 4.50 & 5.3\% \\
\bottomrule
\end{tabular}
\caption{Power decomposition~\eqref{eq:Ptot} at $R/M=150$ ($x=6$) for the benchmark~\eqref{eq:bench}, co-rotating orbit, at $\beta=1$; powers in units of $\PGW$, eq.~\eqref{eq:PGW}. The derived columns are defined in eqs.~\eqref{eq:ceq}--\eqref{eq:Pmin}. The last row is the counter-rotating orbit.}
\label{tab:decomp}
\end{table}

\begin{figure}[t]
\centering
\safefig{\textwidth}{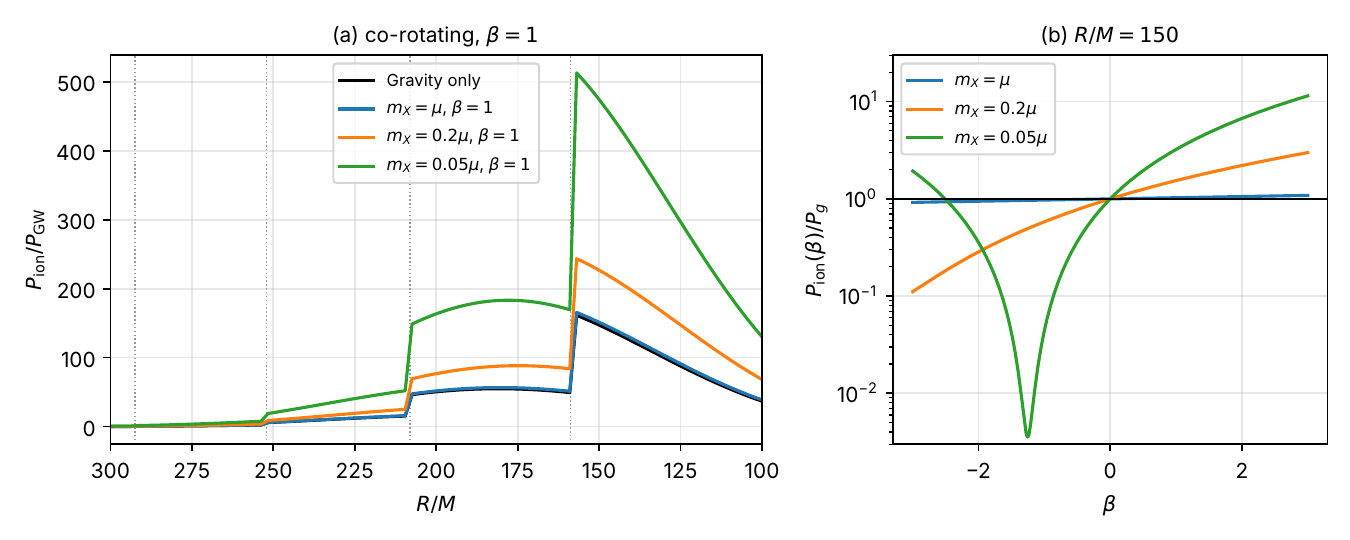}
\caption{(a) Ionization power along the inspiral for the benchmark~\eqref{eq:bench} at $\beta=1$, including gravity, the Yukawa term and their interference. The curve for $m_X=\mu$ lies almost on top of the gravitational one. Dotted lines mark the thresholds~\eqref{eq:thresholds}; each jump is the opening of a new harmonic, eq.~\eqref{eq:zero}. The range shown, $R>R_*^{(1)}$, has $g=1$ closed. (b) $P_{\rm ion}(\beta)/P_g$ at $R/M=150$ from eq.~\eqref{eq:Ptot} and table~\ref{tab:decomp}. 
For $\beta<0$, interference initially suppresses the ionization power, which reaches its minimum at eq.~(3.14).}
\label{fig:power}
\end{figure}

For $m_X=\mu$, the pure Yukawa power equals the gravitational power only at
$|\beta|=\beta_{\rm eq}^{P}=67.95$, whereas a 10\% interference correction requires $|\beta|=3.63$, at which the quadratic term contributes only $0.3\%$ of $P_g$. At $\beta=1$, interference already changes the ionization power by $2.8\%$. Thus, the leading new-force correction at
weak coupling comes from interference rather than the pure Yukawa power.

\subsection{Mediator-mass dependence and the sign of \texorpdfstring{$\beta$}{beta}}

\begin{figure}[t]
\centering
\safefig{0.72\textwidth}{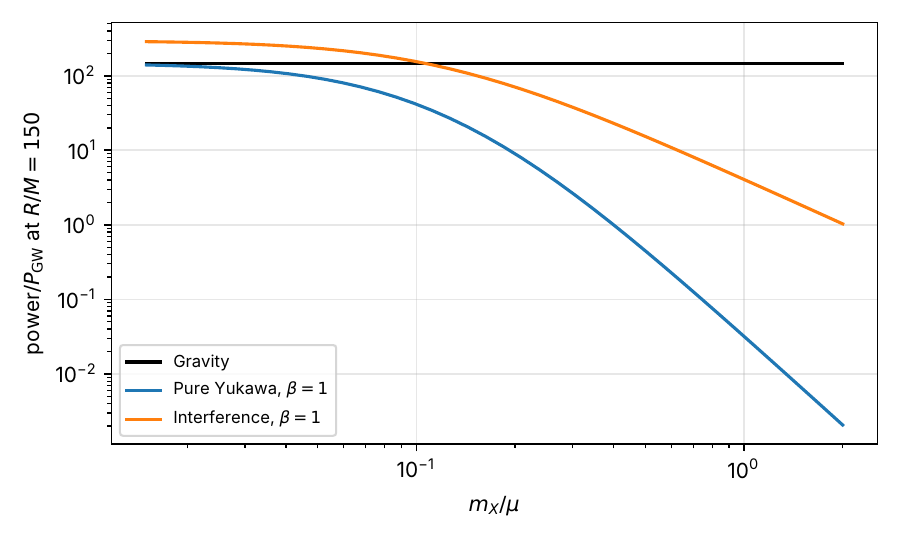}
\caption{Gravity, pure Yukawa and gravity--Yukawa interference at $R/M=150$ and $\beta=1$ as functions of the mediator mass. The interference term is linear in $\beta$ and is the larger of the two new-force terms for all masses shown.}
\label{fig:decomp}
\end{figure}

Figure~\ref{fig:decomp} shows the mediator-mass dependence of the power decomposition at $R/M=150$. Since $P_{\rm int}>0$ throughout the scan, an attractive Yukawa interaction enhances ionization, while a repulsive interaction initially suppresses it. For $\beta<0$ and $|\beta|\ll P_{\rm int}/P_X$,
\begin{equation}
P_{\rm ion}\simeq P_g-|\beta|P_{\rm int}.
\label{eq:repulsion_linear}
\end{equation}
The power reaches its minimum at $\beta_*=-c_P\beta_{\rm eq}^{P}$, as shown in table~\ref{tab:decomp}.

The stronger suppression for lighter mediators follows from the increasing alignment of the gravitational and Yukawa amplitudes, with $c_P=0.937,\ 0.970,\ 0.998$. In the long-range limit, $c_P\to1$ when the $g=1$ channel is closed. When the dipole channel opens, the corrected gravitational
and Yukawa dipoles have different radial profiles.
Their reduced correlation weakens the destructive
interference found at $x=6$, where the dipole channel is closed. The interference has the opposite sign in the $211\to200$ channel discussed in section~\ref{sec:bb}.

The Yukawa contribution is also stronger for counter-rotating orbits. At $m_X=0.2\mu$, $P_X/P_{\rm GW}=27.9$ and $P_{\rm int}/P_{\rm GW}=125.7$, compared with $9.0$ and $70.6$ for the co-rotating orbit, while the gravitational power changes little (table~\ref{tab:decomp}).

The mass scan uses the instantaneous multipole expansion of the Yukawa potential in eq.~(3.4). At the lowest mediator mass, $m_X=0.015\mu$, the condition for on-shell mediator emission, $g\Omega>m_X$, requires $g\geq6$. These harmonics contribute only $0.9\%$ of the calculated $P_X$. This treatment is adequate for the scan, although a retarded propagator is needed for precision near its lower-mass edge (appendix~D.2).

\subsection{Cloud depletion}
\label{sec:drag}
Even when the ionization power is large compared with the gravitational-wave power, the cloud can continue to grow if superradiant replenishment exceeds ionization losses. If $\Gamma_{211}$ is the superradiant amplitude growth rate, the cloud stops growing when
\begin{equation}
  \Gamma_N=\Gamma_{N,g}+\beta\,\Gamma_{N,{\rm int}}+\beta^2\Gamma_{N,X}>2\Gamma_{211}.
  \label{eq:term}
\end{equation}
Here, $\Gamma_N$ is the total occupation-number loss rate due to continuum ionization. The subscripts $g$, $\mathrm{int}$, and $X$ denote the gravitational, gravity--Yukawa interference, and pure Yukawa contributions, respectively, as defined in section 3.2.
Since the occupation number is proportional to the squared amplitude, its growth rate is $2\Gamma_{211}$.

We take $M\Gamma_{211}=2.328\times10^{-9}$, i.e.\ $2M\Gamma_{211}/(q^2\alpha^3)=0.582$ in the units of eq.~\eqref{eq:A4}. This is the value used for all termination couplings in this paper. Because $\Gamma_N\propto q^2$, the required couplings depend on $q$ and $\Gamma_{211}$ only through $\Gamma_{211}/q^2$, and the pure Yukawa values scale as $\Gamma_{211}^{1/2}$.

The criterion~\eqref{eq:term} applies when the companion is present while the cloud is still growing, as in ref.~\cite{TWZ2022}. By contrast, section~\ref{sec:observables} considers an already saturated cloud with $M_c/M=10^{-2}$, for which the black-hole spin has reached the superradiant threshold and $\Gamma_{211}\simeq0$. In this case, the relevant comparison is between the depletion time $\Gamma_N^{-1}$ and the inspiral time.

\begin{figure}[t]
\centering
\safefig{0.75\textwidth}{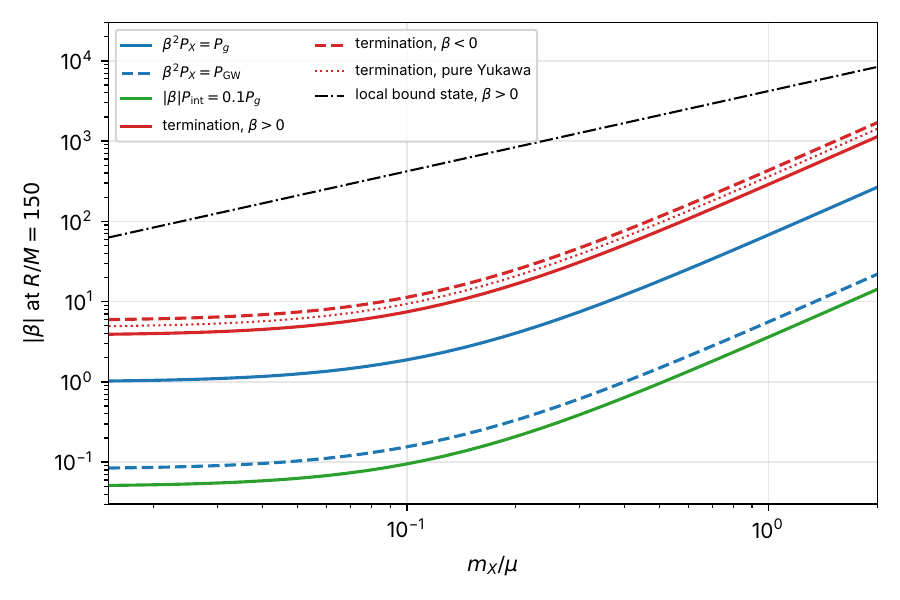}
\caption{Comparison of the characteristic coupling scales associated with different interaction effects at $R/M=150$ as functions of the mediator mass. Blue: pure Yukawa power equal to gravity (solid) or to $\PGW$ (dashed). Green: $10\%$ interference correction. Red: termination of cloud growth, eq.~\eqref{eq:term}, with all terms for $\beta>0$ (solid) and $\beta<0$ (dashed), and with the Yukawa term alone (dotted). Dash-dotted: onset $|\beta|q\alpha\mu/m_X\simeq0.84$ of a bound level centered on the companion for an attractive force, eq.~\eqref{eq:local_yukawa_bound}.}
\label{fig:betamed}
\end{figure}

For a growing cloud, gravity alone gives $\Gamma_{N,g}=0.0251q^2\alpha^3/M$ at $x=6$, only 4.3\% of the rate required by eq. (4.5). Gravitational ionization therefore cannot stop superradiance in this benchmark. Solving eq.~\eqref{eq:term} for $\beta$ gives the couplings in table~\ref{tab:term}. Keeping only the Yukawa term $\beta^2\Gamma_{N,X}$ gives the middle column. Including gravity and interference lowers the required $|\beta|$ by about $20\%$ for $\beta>0$, but raises it by a similar amount for $\beta<0$. The pure Yukawa estimate therefore provides neither an upper nor a lower bound on the termination coupling. Figure~\ref{fig:betamed} shows these couplings and the scales of section~\ref{sec:benchmark} as functions of $m_X$.

\begin{table}[t]
\centering
\begin{tabular}{cccc}
\toprule
$m_X/\mu$ & pure Yukawa & all terms, $\beta>0$ & all terms, $\beta<0$\\
\midrule
1.00 & 361.3 & 288.9 & 432.3\\
0.20 & 20.77 & 16.5 & 25.0\\
0.05 & 6.14 & 4.9 & 7.4\\
\bottomrule
\end{tabular}
\caption{Couplings at which ionization stops the growth of the $\ket{211}$ cloud, eq.~\eqref{eq:term}, at $R/M=150$ with $M\Gamma_{211}=2.328\times10^{-9}$. ``Pure Yukawa'' keeps only $\beta^2\Gamma_{N,X}$.}
\label{tab:term}
\end{table}

Comparing table~\ref{tab:term} with section~\ref{sec:benchmark}, a $10\%$ change of the orbital drag needs a coupling about two orders of magnitude below the one that stops the growth of the cloud.

\section{Continuum ionization versus the \texorpdfstring{$211\to200$}{211->200} channel}
\label{sec:bb}
Continuum ionization and off-resonant $211\to200$ mixing both deplete the cloud but have opposite orbital backreactions. Ionization removes energy and angular momentum from the orbit, whereas the $m=1\to0$ transition transfers them to the orbit before the $|200\rangle$ state is absorbed by the black hole. We compare the two channels first for the Yukawa interaction alone and then including gravity and interference.

\subsection{Pure Yukawa}
\label{sec:bb-pure}
The spherically symmetric $|n00\rangle$ states are strongly absorptive~\cite{TWZ2022}. The $|200\rangle$ level belongs to this class. The dipole component of the companion's Yukawa potential can mix it with the initial $|211\rangle$ cloud, providing a channel for cloud depletion. Only the dipole ($L=1$) connects $\ell_b=1$ to $\ell_f=0$. Since the final state has $m_f=0$, the transition is driven by the $M=-1$ harmonic, which transfers energy $\Omega$ from the cloud to the orbit. The angular factor is
\[
\frac{4\pi}{3}Y^*_{1,-1}\left(\frac{\pi}{2},0\right)
\int Y^*_{00}Y_{1,-1}Y_{11}\,d\Omega
=-\frac{1}{\sqrt{6}},
\]
so the matrix element is
\begin{equation}
\eta_{200,211}
=\frac{q\mu\alpha^2}{\sqrt{6}}
\left[I_1^g+\beta I_1^X\right],
\qquad
I_1=\int_0^\infty d\rho\,\rho^2
\widetilde{\mathcal R}_{20}(\rho)
\widetilde{\mathcal R}_{21}(\rho)
K_1(\rho,x).
\label{eq:mixing_amplitude}
\end{equation}
Here, $K_1$ is the radial function associated with the $L=1$ component of either the gravitational or Yukawa potential.

In the rotating frame, the detuning is
\[
\Delta=\epsilon_{200}-\epsilon_{211}+\Omega,
\]
including the gravitational diagonal shifts induced by the companion. The absorption rate of
$|200\rangle$ is $|\Gamma_{200}|$.
When
\[
|\eta_{200,211}|\ll|\Delta-i\Gamma_{200}|,
\]
the $|200\rangle$ state can be adiabatically eliminated, giving the occupation-number loss rate
\begin{equation}
\Gamma_N^{211\to200}
=\frac{2|\eta_{200,211}|^2|\Gamma_{200}|}
{\Delta^2+\Gamma_{200}^2}.
\label{eq:mixing_rate}
\end{equation}

For the pure Yukawa contribution, we keep only the Yukawa amplitude in eq.~\eqref{eq:mixing_amplitude} and require $\Gamma_N^{211\to200}>2\Gamma_{211}$, using the same $\Gamma_{211}$ as in section~\ref{sec:drag}. At $m_X=\mu$, the minimum coupling magnitude is
\footnote{\label{fn:fixed_detuning}For the numerical estimate, we take $a_*=1$, $M|\Gamma_{200}|=3.2\times10^{-5}$, and $M\Gamma_{211}=2.328\times10^{-9}$ Ref.~\cite{TWZ2022}. The gravitational diagonal shifts induced by the companion
are included in the reference detuning $\Delta$, which is held fixed as $\beta$ varies. We neglect the additional differential diagonal shift due to the static Yukawa potential. With $\Delta$ held fixed and only the Yukawa transition
amplitude retained, the rate is even in $\beta$.}
\begin{equation}
|\beta|_{\min}^{211\to200}
=1.635\times10^3
\qquad\text{at }x=4.676.
\label{eq:beta_bound_termination}
\end{equation}

For comparison, the minimum coupling required for continuum ionization to halt superradiant growth is
\begin{equation}
|\beta|_{\min}^{\rm cont}
=105.57
\qquad\text{at }x=2.454,
\label{eq:beta_cont_termination}
\end{equation}
with the full Yukawa dipole retained (section~\ref{sec:multipole}).  For this benchmark, the $211\to200$ threshold is about 15 times larger in coupling at the respective optimal separations. For the pure Yukawa estimates, both rates scale as
$\beta^2$, with $\Delta$ held fixed as $\beta$ varies in the $211\to200$ channel; their rate coefficients therefore differ by a factor of about 240 (figure 4).

For $m_X=\mu$, the static Yukawa shift of the $|211\rangle$ binding energy remains below $1\%$ at both thresholds. This small shift reflects the screening of the Yukawa potential over most of the cloud. At the $211\to200$ threshold, the mixing parameter remains small. The Born parameter controlling the distortion of the final states, however, is not parametrically small, although it remains below the threshold for a local bound state around the companion for the attractive sign. The $211\to200$ threshold should therefore be regarded as an estimate (Appendix D.1).
  
\begin{figure}[t]
\centering
\safefig{0.75\textwidth}{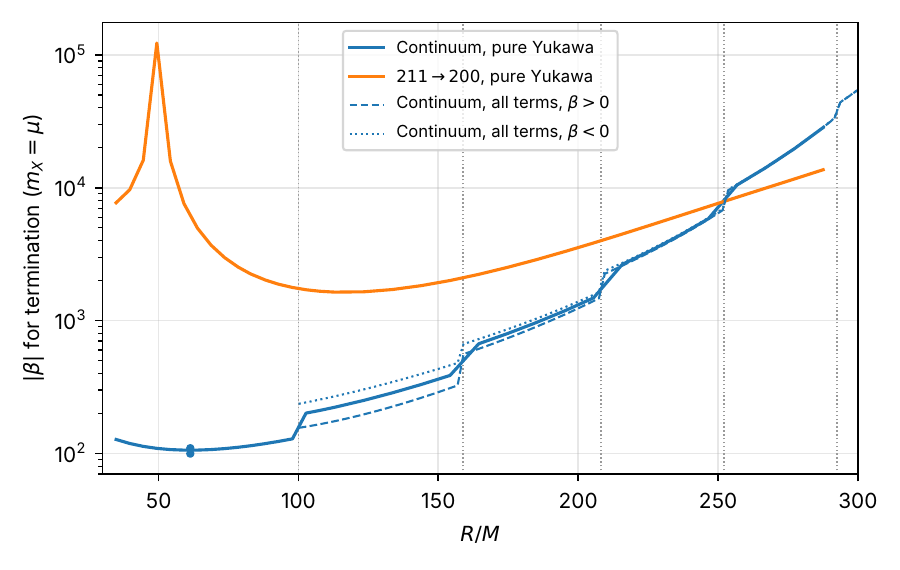}
\caption{Coupling needed to stop the growth of the $\ket{211}$ cloud for $m_X=\mu$, through ionization into the continuum (blue) or through off-resonant $211\to200$ mixing (orange). Solid curves keep only the Yukawa amplitude. Dashed and dotted blue curves include gravity and the interference for $\beta>0$ and $\beta<0$ in the region $R>R_*^{(1)}$; the two points at $R/M=61$ are the full values at the continuum optimum, $|\beta|=99.5$ and $109.4$. The $211\to200$ curve does not include the gravitational dipole amplitude discussed in section~\ref{sec:bb-grav}. Dotted vertical lines mark the thresholds~\eqref{eq:thresholds}.}
\label{fig:cvb}
\end{figure}

\subsection{Gravity and interference in both channels}
\label{sec:bb-grav}
The gravitational and Yukawa amplitudes interfere in both channels because they connect the same initial and final states. For continuum ionization, this interference is already included in section 3.2. Including gravity shifts the pure Yukawa termination coupling only modestly, lowering it for an attractive interaction and raising it for a repulsive one. Gravity alone contributes only a few percent of the rate required to halt superradiance.

For the $211\to200$ channel, the gravitational contribution comes from the corrected dipole in eq.~\eqref{eq:dipole}. This dipole vanishes for $\rho<x$ and survives only outside the orbit. At $x=4.676$, $49.9\%$ of the $|211\rangle$ cloud lies at
$\rho>x$, so the gravitational dipole contributes to the transition. The radial integrals in eq.~\eqref{eq:mixing_amplitude} give
\[
\frac{I_1^g}{I_1^X}=
\begin{cases}
-119, & m_X=\mu,\\
-6.2, & m_X=0.2\mu,\\
-1.5, & m_X=0.05\mu.
\end{cases}
\]
For $m_X=\mu$, the gravitational amplitude therefore has the same magnitude as a Yukawa amplitude with $|\beta|\simeq119$, but the opposite sign. This follows because the inertial term $-\rho/x^2$ dominates the corrected dipole outside the orbit. Consequently, an attractive interaction ($\beta>0$) interferes destructively in this channel, whereas a repulsive interaction
($\beta<0$) interferes constructively.

At the pure Yukawa threshold $|\beta|=1.635\times10^3$, the gravitational amplitude is about $7.3\%$ of the Yukawa
amplitude. At fixed detuning, this changes the rate by approximately $-14\%$ for $\beta>0$ and $+15\%$ for $\beta<0$, shifting the estimated thresholds to $1.75\times10^3$ and
$1.52\times10^3$, respectively. These shifted thresholds should therefore be interpreted within the fixed-detuning approximation described in footnote~\ref{fn:fixed_detuning}. Unlike continuum ionization, the $211\to200$ channel has no kinematic opening threshold: it is an off-resonant mixing with a bound state. Its rate can nevertheless vanish at a particular coupling if the two amplitudes cancel. Equation~\eqref{eq:beta_bound_termination} therefore gives the coupling required to halt superradiant growth, not the coupling at which the transition first becomes allowed.

At a fixed separation, the kinematic factor is common to the two interactions: $\mu/k_*$ for continuum ionization and the Lorentzian factor in eq.~\eqref{eq:mixing_rate} for $211\to200$. Their relative strengths and interference are therefore characterized by $\beta_{\rm eq}^{N}$ and $c_N$. For the bound-state channel, this comparison uses the fixed-detuning approximation described in section~\ref{sec:bb-pure}.
\begin{table}[!htbp]
\centering
\begin{tabular}{lcccc}
\toprule
 & \multicolumn{2}{c}{$x=4.676$ ($R/M=116.9$)} & \multicolumn{2}{c}{$x=6$ ($R/M=150$)}\\
\cmidrule(lr){2-3}\cmidrule(lr){4-5}
 & $m_X=\mu$ & $m_X=0.2\mu$ & $m_X=\mu$ & $m_X=0.2\mu$\\
\midrule
Continuum: $\beta_{\rm eq}^{N}$ ($c_N$) & 51.1 (0.97) & 3.29 (0.99) & 75.0 (0.96) & 4.31 (0.98)\\
$211\to200$: $\beta_{\rm eq}^{N}$ ($c_N$) & 119 ($-1$) & 6.2 ($-1$) & 80 ($-1$) & 3.5 ($-1$)\\
\bottomrule
\end{tabular}
\caption{Relative strengths and interference of gravitational and Yukawa contributions in the continuum and $211\to200$ channels at two orbital separations. Here $\beta_{\rm eq}^{N}=(\Gamma_{N,g}/\Gamma_{N,X})^{1/2}$ and $c_N=\Gamma_{N,\rm int}/ (2\sqrt{\Gamma_{N,g}\Gamma_{N,X}})$ are defined using occupation-number rates.}
\label{tab:2x2}
\end{table}

At both separations, the two channels have comparable $\beta_{\rm eq}^{N}$, indicating similar relative strengths of the Yukawa and gravitational contributions. The factor of about 240 between the pure Yukawa rate coefficients found in section~\ref{sec:bb-pure} therefore reflects the different efficiencies of the two channels, rather than a different relative strength of the Yukawa interaction. The same channel dependence appears in the gravitational rates. Table~\ref{tab:2x2} also shows that the interference has opposite signs in the two channels: $c_N>0$ for continuum ionization and $c_N=-1$ for $211\to200$. Thus, an attractive interaction enhances continuum ionization but suppresses the $211\to200$ transition, while a repulsive interaction has the opposite effects.

Finally, gravity alone cannot stop superradiant growth in either channel for this benchmark. In the $211\to200$ channel, its amplitude corresponds to $|\beta|\simeq119$, contributing only $(119/1635)^2\simeq0.5\%$ of the required depletion rate at $x=4.676$. The Yukawa interaction therefore provides most of the depletion, with continuum ionization reaching the termination condition at a substantially smaller coupling.

\section{Orbital evolution and the phase correction to gravitational waves}
\label{sec:observables}
The continuum-ionization power derived in sections~\ref{sec:yukawa} and~\ref{sec:results} modifies the orbital energy loss of the binary and therefore its inspiral and gravitational-wave phase.\footnote{The orbital evolution considered here
does not include the sequence of resonances associated with highly excited bound states
close to the continuum threshold, which can accumulate below the continuum threshold and smooth the features associated with the opening of continuum channels \cite{tomaselli2025smooth,tomaselli2024resonant}.} In this section, we use energy balance to relate the ionization power to the orbital evolution and derive the corresponding correction to the gravitational-wave phase.

\subsection{Setup and orbital range}
\label{sec:obs-input}
We retain the phenomenological interaction introduced
in section~\ref{sec:intro}, with $\beta>0$ ($\beta<0$) corresponding
to an attractive (repulsive) Yukawa interaction. As
before, the central black hole is assumed to carry no
Yukawa charge.

We restrict the analysis to the orbital band
\begin{equation}
R_*^{(1)}<R\leq 300M,
\qquad
4<x\leq 12
\label{eq:orbital_band}
\end{equation}
In this range, the $g=1$ continuum channel remains closed, so only harmonics with $g\geq 2$ contribute to ionization. Since the angular selection rules require $L\geq g$, the dipole term with $L=1$ does not contribute. The analysis is therefore independent of the
dipole contribution, including the inertial correction in the black-hole frame discussed
in sections 3.1 and 5.

\subsection{Orbital evolution}
\label{sec:obs-evolution}
For a quasi-circular orbit, the Newtonian binding energy is
\begin{equation}
E_{\rm orb}
=-\frac{qM^2}{2R}
=-\frac{qM\alpha^2}{2x}.
\label{eq:orbital_energy}
\end{equation}
Using energy balance,
\begin{equation}
\dot E_{\rm orb}
=-(P_{\rm GW}+P_{\rm ion}),
\label{eq:energy_balance}
\end{equation}
where $P_{\rm GW}$ is the vacuum gravitational-wave luminosity and $P_{\rm ion}$ is the total ionization power, we obtain
\begin{equation}
\dot x
=-\frac{2x^2}{qM\alpha^2}
(P_{\rm GW}+P_{\rm ion}),
\qquad
\dot\Omega
=\frac{3x\Omega}{qM\alpha^2}
(P_{\rm GW}+P_{\rm ion}).
\label{eq:orbital_evolution}
\end{equation}

At fixed orbital frequency and cloud mass, the chirp rate is therefore proportional to the total dissipated power. Using the interference at the amplitude level in eq.~(3.10), the ionization power takes the form
\begin{equation}
P_{\rm ion}=P_g\left[
1+2c_P(x)\frac{\beta}{\beta_{\rm eq}^{P}(x)}
+\left(\frac{\beta}{\beta_{\rm eq}^{P}(x)}\right)^2
\right].
\label{eq:Pion_obs}
\end{equation}
The overlap coefficient $c_P$ is positive throughout the orbital band considered. For weak coupling at fixed $M_c$,
\begin{equation}
\left.
\frac{\partial\ln\dot\Omega}{\partial\beta}
\right|_{\Omega,M_c,\beta=0}
=
\frac{P_{\rm int}^{(1)}}{P_{\rm GW}+P_g}
=
\frac{2c_P}{\beta_{\rm eq}^{P}}
\frac{P_g}{P_{\rm GW}+P_g}.
\label{eq:chirp_response}
\end{equation}
Figure 5 shows this response. The changes at the harmonic thresholds reflect the opening of additional ionization channels.

The same ionization transitions deplete the cloud according to
\begin{equation}
\frac{\dot M_c}{M_c}=-\Gamma_N,
\qquad
M\Gamma_N=q^2\alpha^3
\sum_{\ell_f,g}
\frac{
\left|A^{(g)}_{g,\ell_fm_f}
+\beta A^{(g)}_{X,\ell_fm_f}\right|^2
}{\kappa_g},
\label{eq:cloud_depletion}
\end{equation}
where $A_g^{(g)}$ and $A_X^{(g)}$ are the gravitational and Yukawa amplitudes defined in appendix~\ref{app:matrix}. Since $P_{\rm ion}\propto M_c$, the orbital evolution and cloud depletion equations are integrated together.

As discussed in section~\ref{sec:drag}, we consider a cloud that has already reached superradiant saturation, with $M_c/M=10^{-2}$ and $\Gamma_{211}\simeq0$. Further superradiant growth is therefore neglected during the inspiral.

\begin{figure}[t]
  \centering
  \safefig{0.72\textwidth}{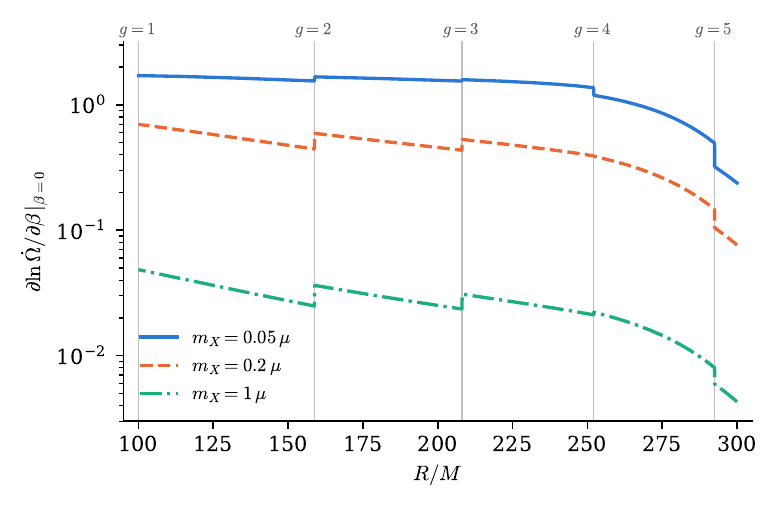}
  \caption{
Linear response of the chirp rate to the Yukawa
coupling,
$\left.\partial\ln\dot\Omega/\partial\beta\right|_{\Omega,M_c,\beta=0}$,
along the orbital band~\eqref{eq:orbital_band} for the benchmark of
section~\ref{sec:benchmark}. The curves correspond to
$m_X/\mu=0.05,\ 0.2,\ 1$, and the vertical lines mark
the thresholds $R_*^{(g)}$ at which the corresponding
continuum harmonics open. The response is positive
throughout the band, so the leading correction
increases (decreases) the chirp rate for $\beta>0$
($\beta<0$).
}
  \label{fig:chirp-response}
\end{figure}

\subsection{The phase correction to gravitational waves}
\label{sec:obs-dephasing}
For the leading quadrupole harmonic, the accumulated
gravitational-wave phase is
\begin{equation}
\Phi_{\rm GW}=2\int\Omega\,dt.
\label{eq:phase_def}
\end{equation}
Using the orbital evolution in eq.~(6.4),
we obtain
\begin{equation}
\Phi_{\rm GW}(\beta)
=q\sqrt{1+q}\alpha^5
\int_{x_{\rm in}}^{x_{\rm out}}
\frac{dx}{x^{7/2}[P_{\rm GW}+P_{\rm ion}]},
\label{eq:phase_integral}
\end{equation}
where the integration limits correspond to fixed orbital frequencies. The cloud mass $M_c(x;\beta)$ is obtained by integrating eqs.~\eqref{eq:orbital_evolution} and~\eqref{eq:cloud_depletion} simultaneously.

We define the phase correction relative to the case with gravitational ionization alone as
\begin{equation}
\delta\Phi(\beta)
\equiv\Phi_{\rm GW}(\beta)-\Phi_{\rm GW}(0)
=\Phi_1\beta+\Phi_2\beta^2+O(\beta^3).
\label{eq:phase_expansion}
\end{equation}
Here, $\Phi_1$ and $\Phi_2$ are the first two Taylor coefficients of the accumulated phase at $\beta=0$. The linear term captures the leading interference effect, including its influence on cloud depletion, while the quadratic term also receives contributions from the nonlinear dependence of the phase on the total ionization power.

The phase calculation retains the dissipative effect of the Yukawa interaction. The static Yukawa interaction also modifies the conservative orbit and shifts the bound-state energy, thereby changing the relation between orbital frequency and orbital separation, as well as the ionization thresholds. These corrections are estimated in appendix~\ref{app:static} and are subleading in the parameter range considered.

\subsection{Frequency dependence of the ionization signal}
\label{sec:obs-degeneracy}
At fixed separation, the Yukawa correction in eq.~\eqref{eq:Pion_obs} can largely be mimicked by changing the cloud mass, since $P_g\propto M_c$.
Over the band~(6.1), $c_P\simeq1$ and $\beta_{\rm eq}^{P}$ varies slowly, making most of the smooth phase correction degenerate with $M_c$. Constant offsets in phase and time can also be absorbed into the
corresponding reference parameters.

The frequency dependence of $c_P/\beta_{\rm eq}^{P}$ is most pronounced near the harmonic thresholds. When the $h$-th harmonic opens at $R_*^{(h)}$, its additional contribution to the ionization power changes the chirp rate.
From eq.~\eqref{eq:orbital_evolution}, $\dot{\Omega}$ is proportional to $P_{\rm GW}+P_{\rm ion}$ at fixed orbital frequency. We therefore define the local fractional
change in the chirp rate as
\begin{equation}
s_h\equiv
\left.\frac{\Delta\dot{\Omega}}{\dot{\Omega}}
\right|_{R=R_*^{(h)}}
=
\left.
\frac{\Delta P_h}{P_{\rm GW}+P_{\rm ion}}
\right|_{R\rightarrow R_*^{(h)+}},
\label{eq:chirp_step}
\end{equation}
where $\Delta P_h$ is the finite contribution of the newly opened harmonic, as given by eq.~\eqref{eq:zero}. The thresholds occur at the fixed gravitational-wave frequencies
\[
f_*^{(h)}=\frac{\mu\alpha^2}{8\pi h}.
\]
Since the ratio of Yukawa to gravitational amplitudes for the newly opened harmonic
generally differs from that for the harmonics already open, the step height depends on $\beta$. When ionization dominates the energy loss, the fractional step is only weakly sensitive to $M_c$.

The steps are discontinuous only in the Markov limit. A finite chirp rate spreads each transition over a narrow frequency interval, so the relevant quantity is the change in the chirp rate across the threshold rather than an exactly instantaneous jump. The distinction also requires an accurate gravitational ionization baseline, since an error in $P_g$ that varies with frequency can mimic a change in the power spectrum caused by the Yukawa interaction.

\section{Conclusions}
\label{sec:conclusions}
We have studied the effect of a companion-sourced Yukawa interaction on the ionization of a $|211\rangle$ gravitational atom.

Gravity and the Yukawa interaction induce transitions to the same continuum states,
allowing their amplitudes to interfere. The interference provides the leading correction at weak coupling, even when the pure Yukawa power is much smaller than the gravitational contribution. At larger coupling, the pure Yukawa contribution becomes increasingly important and can compete with the gravitational contribution, particularly for lighter mediators. The effect also depends on the mediator range and the sign of $\beta$. When the $g=1$ continuum channel is closed, the gravitational and Yukawa amplitudes are nearly parallel, allowing a repulsive interaction to strongly suppress ionization. This suppression becomes less effective once this channel opens.

Continuum ionization and off-resonant $211\to200$ mixing provide distinct channels for cloud depletion and have opposite effects on the orbital energy. In the benchmark considered here,
gravitational ionization alone supplies only a few percent of the rate required to halt superradiant growth, while the continuum channel reaches the depletion condition at a substantially smaller Yukawa coupling than the $211\to200$ channel. Moreover, the interference has opposite signs in the two channels: an attractive Yukawa interaction enhances continuum ionization but suppresses $211\to200$ mixing, with the reverse behavior for a repulsive interaction. The continuum channel is therefore essential when assessing cloud depletion by a binary companion.

The ionization also affects the coupled evolution of the orbit and the cloud. An attractive Yukawa interaction ($\beta>0$) accelerates the inspiral, whereas a repulsive one ($\beta<0$) slows it over the orbital band considered, with both effects occurring at weak coupling. Although the smooth accumulated phase correction is largely degenerate with the cloud mass, the frequency dependence of the continuum-ionization power at successive harmonic thresholds produces local changes in the chirp rate within the model considered here. We characterize this response using $s_h$ in eq.~(6.11). When continuum
ionization dominates the energy loss, this diagnostic is only weakly sensitive to $M_c$.
Whether it remains distinguishable in the full gravitational-wave waveform requires including bound-state excitations and their backreaction on the orbit and the cloud. 

Our treatment assumes a point companion with no Yukawa charge on the central black hole, a cloud that remains in the $|211\rangle$ state, and a quasi-circular, equatorial orbit. It also uses non-relativistic continuum states of the 1/r gravitational potential and an instantaneous form of the Yukawa potential. Further work will incorporate bound-state resonances close to the continuum threshold into the coupled orbital and cloud evolution, allowing a more complete analysis of the associated gravitational-wave signatures. The present framework can also be extended to other source configurations and more general orbits, while accounting for the prior evolution of the cloud and relativistic effects.

\section*{Acknowledgments}
We are especially grateful to Professor Yue Zhao for helpful discussions. JY also thanks Kaiyuan Fan, Shengyi Liu, Yonghao Liu, Yanjiao Ma, Xin Meng, Haoxuan Sun,
Xiangwei Wang, Xingkai Zhang, and Yunke Zhao for their discussions and support during the summer research period. We acknowledge ARC (Agent Research Copilot), described in Ref.~\cite{ma2026arc} (GitHub: \url{https://github.com/tririver/arc}), for providing research ideas and assistance with some calculations. We also acknowledge the use of ChatGPT and Claude for assistance with manuscript organization and checking selected calculations.

\appendix
\section{Angular and radial matrix elements}
\label{app:matrix}

For an equatorial source the angular coefficient is the Gaunt integral
\begin{align}
  \mathcal G_{\ell_fm_f;Lg;\ell_im_i}&=\int\dd\Omega\,Y^*_{\ell_fm_f}Y_{Lg}Y_{\ell_im_i}\nonumber\\
  &=(-1)^{m_f}\sqrt{\frac{(2\ell_f+1)(2L+1)(2\ell_i+1)}{4\pi}}
  \begin{pmatrix}\ell_f&L&\ell_i\\0&0&0\end{pmatrix}
  \begin{pmatrix}\ell_f&L&\ell_i\\-m_f&g&m_i\end{pmatrix}.
  \label{eq:A1}
\end{align}
It enforces
\begin{equation}
  m_f=m_i+g,\qquad|\ell_f-\ell_i|\le L\le\ell_f+\ell_i,\qquad \ell_f+L+\ell_i\ \text{even}.
  \label{eq:A2}
\end{equation}
The source factor $Y^*_{Lg}(\pi/2,0)$ vanishes for odd $L+g$, which with eq.~\eqref{eq:A2} gives $\ell_f+m_f$ even: the source and $\ket{211}$ are even under $z\to-z$, so only even final states are reached. In addition $L\ge g$. For each open harmonic the dimensionless matrix element is
\begin{equation}
  \mathcal A^{(g)}_{\ell_fm_f}=\sum_L\frac{4\pi}{2L+1}\,Y^*_{Lg}\!\left(\frac\pi2,0\right)\mathcal G_{\ell_fm_f;Lg;\ell_im_i}\,I_{\ell_fL},
  \label{eq:A3}
\end{equation}
with $I_{\ell_fL}=I^g_{\ell_fL}$ for gravity and $I^X_{\ell_fL}$ for the Yukawa term, eq.~\eqref{eq:IX}. With eq.~\eqref{eq:etaX}, $\mu|\eta^{(g)}|^2/k_*=(q^2\alpha^3/M)|\mathcal A^{(g)}|^2/\kappa_g$, and the rate and power reduce to
\begin{align}
  M\,\Gamma_N&=q^2\alpha^3\sum_{\ell_f,g}\frac{\bigl|\mathcal A^{(g)}_{\ell_fm_f}\bigr|^2}{\kappa_g},
  \label{eq:A4}\\
  P_{\rm ion}&=\frac{\Mc}{M}\,q^2\alpha^5\sum_{\ell_f,g}g\,\omega(x)\,\frac{\bigl|\mathcal A^{(g)}_{\ell_fm_f}\bigr|^2}{\kappa_g}.
  \label{eq:A5}
\end{align}
For the total amplitude, $\mathcal A\to\mathcal A_g+\beta\mathcal A_X$, which gives eq.~\eqref{eq:Pparts}. For example, at $x=6$ the leading channel is $g=2$, $\ell_f=m_f=3$, with $L=2,4$; the angular factors are $3\sqrt{210}/140$ and $\sqrt{210}/504$, and this single channel gives $70\%$ of $P_g$.

The zero-energy limit~\eqref{eq:zero} follows from the Kummer form
\begin{equation}
  F_\ell(\eta_{\rm C},z)=C_\ell(\eta_{\rm C})\,z^{\ell+1}e^{-iz}M(\ell+1-i\eta_{\rm C},2\ell+2,2iz).
\end{equation}
For $\kappa\to0$ at fixed $\rho$, the confluent function tends to
\begin{equation*}
\Gamma(2\ell+2)(2\rho)^{-(2\ell+1)/2}J_{2\ell+1}(2\sqrt{2\rho}),
\end{equation*}
The Coulomb coefficient satisfies
\begin{equation*}
C_\ell\to\frac{2^\ell\sqrt{2\pi}\,\kappa^{-\ell-1/2}}{\Gamma(2\ell+2)}.
\end{equation*}
Consequently, $F_\ell\to\sqrt{\pi\kappa\rho}\,J_{2\ell+1}(2\sqrt{2\rho})$
and $\Rt_{\kappa\ell}\to\sqrt{4\pi\kappa/\rho}\,J_{2\ell+1}(2\sqrt{2\rho})$.

The same zero-energy limit also matches the highly excited bound-state spectrum. For fixed $\ell$, the normalized radial functions satisfy
\begin{equation}
\lim_{n\to\infty}
\sqrt{2\pi n^3}\,\widetilde R_{n\ell}(\rho)
=
\lim_{\kappa\to0^+}
\frac{\widetilde R_{\kappa\ell}(\rho)}{\sqrt{\kappa}}
=
\sqrt{\frac{4\pi}{\rho}}\,
J_{2\ell+1}(2\sqrt{2\rho}).
\end{equation}
Let $\mathcal M_n$ and $\mathcal M_\kappa$ denote the matrix elements of the same harmonic perturbation between $|211\rangle$ and the corresponding normalized
bound and continuum states. Since $\epsilon_n=-1/(2n^2)$ gives $dn/d\epsilon_n=n^3$, their spectral weights obey
\begin{equation}
2\pi n^3|\mathcal M_n|^2
\longrightarrow
\lim_{\kappa\to0^+}
\frac{|\mathcal M_\kappa|^2}{\kappa}.
\end{equation}
This relation holds separately for the gravitational, Yukawa, and interference contributions when the same total perturbation, $V_g+V_X$, is used in both matrix elements. Each transition extracts the same orbital energy $g\Omega$.

\section{Numerical method and validation}
\label{app:reproduction}

\subsection{Continuum wave functions and normalization}

The bound radial functions are evaluated in their normalized Laguerre form. For the continuum, we solve the radial equation for the attractive $1/r$ potential in $\rho$. Writing
\begin{equation}
\widetilde R_{\kappa\ell}(\rho)
=2N_\ell\rho^\ell y_\ell(\rho),
\end{equation}
the function $y_\ell$ satisfies
\begin{equation}
y_\ell''+\frac{2(\ell+1)}{\rho}y_\ell'
+\left(\kappa^2+\frac{2}{\rho}\right)y_\ell=0.
\end{equation}
We integrate from the regular Frobenius initial conditions
$y_\ell(0)=1$ and $y_\ell'(0)=-1/(\ell+1)$.
Matching to
$F_\ell\sim C_\ell(-1/\kappa)(\kappa\rho)^{\ell+1}$
at the origin fixes
\begin{equation}
N_\ell=C_\ell(-1/\kappa)\kappa^{\ell+1},
\qquad
C_\ell(\eta_C)=
\frac{2^\ell e^{-\pi\eta_C/2}
|\Gamma(\ell+1+i\eta_C)|}
{\Gamma(2\ell+2)}.
\end{equation}
In the limit $\kappa\to0$, $N_\ell\propto\kappa^{1/2}$, in agreement with eq.~\eqref{eq:zero}. We use the zero-energy expression directly for $\kappa<8\times10^{-3}$.

\subsection{Numerical convergence and cross-checks}

The radial integrations are split at $\rho=x$. The radial functions in the Yukawa multipole expansion are evaluated using exponentially scaled modified Bessel functions, and the angular coefficients are computed from Wigner $3j$ symbols.

The production runs use $\ell_f\leq12$. Increasing the cutoff to $\ell_f\leq14$ changes the total power by $1.8\times10^{-6}$ for $m_X=\mu$ and $2.8\times10^{-6}$ for $m_X=0.2\mu$, both at $x=6$,
and by $1.3\times10^{-5}$ near the gravitational peak. Individual subleading components converge to $10^{-4}$ or better. Increasing both cutoffs to $\ell_f\leq14$ and $g\leq18$ at $x=6$ changes $P_g$ by $1.6\times10^{-6}$ and $P_{\min}/P_g=1-c_P^2$ by at most $2\times10^{-4}$ of its value. The cancellation in eq.~\eqref{eq:Pparts} is therefore not a truncation artefact.

We also performed independent cross-checks. A direct evaluation using \texttt{mpmath}, Gaunt coefficients and Simpson integration up to $\rho=90$, with $g\leq14$ and $\ell_f\leq12$, reproduces the entries of table~\ref{tab:decomp} to four or five significant figures and the thresholds in eq.~\eqref{eq:beta_cont_termination}. A second implementation using a Numerov integration with grid spacing $0.005$ agrees with the first to nine digits for table~\ref{tab:decomp} and was used for
figures~\ref{fig:power}--\ref{fig:betamed} and section~\ref{sec:observables}. The public GrAB implementation of ref.~\cite{TSB2023} provides an additional check of the continuum normalization, the factor $\mu/k_*$ in the rate, the $4\pi$ multipole convention and the partial-wave convergence.

\subsection{Reproduction of the gravitational benchmark}

\begin{figure}[t]
\centering
\safefig{0.72\textwidth}{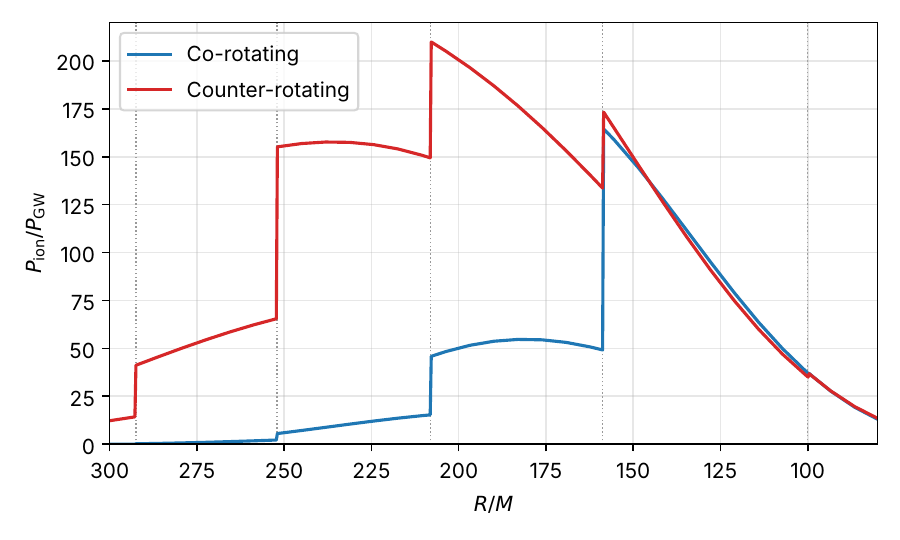}
\caption{Reproduction of the ionization-to-GW power ratio of figure~7 of ref.~\cite{BBST2022}. Dotted lines mark the thresholds~\eqref{eq:thresholds}. The separation axis runs in the direction of the inspiral.}
\label{fig:repro}
\end{figure}

Figure~\ref{fig:repro} and table~\ref{tab:repro} show our reproduction of the gravitational ionization power in figure~7 of ref.~\cite{BBST2022}, for the benchmark in eq.~\eqref{eq:bench} and for co- and counter-rotating orbits. The values in table 4 use $\ell_f\leq22$. As in the original calculation, the dipole is omitted. The thresholds agree with eq.~\eqref{eq:thresholds}. The largest values on our grid are
$P_{\rm ion}/P_{\rm GW}=164.7$ for co-rotation and $210.1$ for counter-rotation, confirming that ionization can exceed the vacuum gravitational-wave luminosity by about two orders of magnitude.

\begin{table}[t]
\centering
\caption{Selected points of the reproduction, $P_{\rm ion}/\PGW$.}
\label{tab:repro}
\begin{tabular}{ccc}
\toprule
$R/M$ & co-rotating & counter-rotating\\
\midrule
100 & 37.313 & 37.189\\
125 & 87.898 & 83.672\\
150 & 147.277 & 149.397\\
200 & 50.469 & 200.714\\
250 & 6.205 & 155.912\\
300 & 0.165 & 12.422\\
\bottomrule
\end{tabular}
\end{table}

\section{Effect of the static \texorpdfstring{$g=0$}{g=0} Yukawa harmonic}
\label{app:static}

The $g=0$ harmonic cannot ionize, since eq.~\eqref{eq:kappag} gives $\kappa_0^2=-1/n^2<0$. It does shift and dress the bound state. The static part of $V_X$ consists of its $M=0$ components; those with odd $L$ vanish because $Y_{L0}(\pi/2,0)=0$ for odd $L$. Its expectation value in $\ket{211}$ is
\begin{gather}
  \delta\epsilon_b=\langle b|V_X^{(0)}|b\rangle=-\beta q\mu\alpha^2\,X(x,\zeta),\nonumber\\
  X(x,\zeta)=\int_0^\infty\dd\rho\,\rho^2\Rt_{21}^2(\rho)\Bigl[K^X_0(\rho,x;\zeta)+\tfrac{1}{10}K^X_2(\rho,x;\zeta)\Bigr],
  \label{eq:C1}
\end{gather}
where $1/10=P_2(0)\langle P_2(\cos\theta)\rangle_{211}$ and higher even multipoles do not contribute, since the $\ket{211}$ density contains only $P_0$ and $P_2$. We measure the size of the static perturbation by
\begin{equation}
  \epsilon_{\rm static}\equiv\frac{|\delta\epsilon_b|}{|\epsilon_b|}=8\,|\beta|\,q\,X(x,\zeta).
  \label{eq:C2}
\end{equation}
Table~\ref{tab:X} gives $X$ at the separations used in the text. The $M=0$ components couple $\ket{211}$ only to states with $m=1$, and $n=2$ contains no other $m=1$ state, so the first-order dressing
\begin{equation}
  |\delta b\rangle=\sum_{i\ne b}|i\rangle\,\frac{\langle i|V_X^{(0)}|b\rangle}{\epsilon_b-\epsilon_i}
  \label{eq:C3}
\end{equation}
has no degenerate denominators; the nearest $m=1$ levels lie at least $\mu\alpha^2(1/8-1/18)$ away. Its size relative to $\ket b$ is of order $\epsilon_{\rm static}$.

\begin{table}[t]
\centering
\caption{Static overlap $X(x,\zeta)$ of eq.~\eqref{eq:C1}. The limit $\zeta\to0$ is the unscreened Newtonian value.}
\label{tab:X}
\begin{tabular}{ccccc}
\toprule
$x$ & $m_X=\mu$ & $m_X=0.2\mu$ & $m_X=0.05\mu$ & $\zeta\to0$\\
\midrule
2.454 & $1.26\times10^{-3}$ & $2.13\times10^{-2}$ & $0.102$ & 0.244\\
4.676 & $5.07\times10^{-4}$ & $1.09\times10^{-2}$ & $0.0673$ & 0.196\\
6 & $2.24\times10^{-4}$ & $5.62\times10^{-3}$ & $0.0464$ & 0.165\\
\bottomrule
\end{tabular}
\end{table}

\subsection{Subleading effects of the static harmonic}

The dressed state $\ket{b'}=\ket b+\ket{\delta b}$ changes the gravitational amplitude to $\eta'_g=\eta_g+\delta\eta_g$ with $\delta\eta_g/\eta_g=O(\epsilon_{\rm static})$, so
\begin{equation}
  \frac{\delta P_g}{P_g}\simeq\frac{2\,{\rm Re}(\eta_g^*\delta\eta_g)}{|\eta_g|^2}=O(\epsilon_{\rm static}),
  \label{eq:C4}
\end{equation}
which is linear in $\beta$. Gravity's own static tide ($L=2$, $M=0$) dresses $\ket b$ at relative order $q$ in the same way. It is larger than the Yukawa dressing for small $|\beta|$ and is neglected here, as in ref.~\cite{BBST2022}.

With the convention~\eqref{eq:etatot}, the Yukawa amplitude is $\beta\eta_X$ and becomes $\beta(\eta_X+\delta\eta_X)$ with $\delta\eta_X/\eta_X=O(\epsilon_{\rm static})$. Hence $\delta P_X/P_X=O(\epsilon_{\rm static})$.

The interference term $2\beta\,{\rm Re}(\eta_g^*\eta_X)$ becomes $2\beta\,{\rm Re}[(\eta_g+\delta\eta_g)^*(\eta_X+\delta\eta_X)]$, and the corrections are again of relative order $\epsilon_{\rm static}$.  Away from threshold, the static-harmonic corrections to the three contributions are all perturbative in $\epsilon_{\rm static}$,
\begin{equation}
  \frac{\delta P_g}{P_g}\sim\frac{\delta P_X}{P_X}\sim\frac{\delta\Pint}{\Pint}\sim\epsilon_{\rm static}.
  \label{eq:C5}
\end{equation}
At the termination thresholds of section~\ref{sec:bb}, $\epsilon_{\rm static}=6.6\times10^{-3}$ ($|\beta|=1635$, $x=4.676$) and $1.1\times10^{-3}$ ($|\beta|=105.6$, $x=2.454$), both for $m_X=\mu$. At $|\beta|=1$ it is below $4\times10^{-4}$ for all three benchmark masses at $x=6$.

Near a threshold the energy shift matters more than the dressing, because it moves the threshold itself. With $\delta\epsilon_b$ included, the opening condition becomes $g\,\omega(x_g)=1/8-\delta\epsilon_b/(\mu\alpha^2)$, and since $\omega\propto x^{-3/2}$,
\begin{equation}
  \frac{\delta x_g}{x_g}=\frac23\,\frac{\delta\epsilon_b}{|\epsilon_b|}=-\frac{16}{3}\,\beta q\,X(x_g,\zeta).
  \label{eq:C6}
\end{equation}
An attractive force binds the cloud more strongly and moves every threshold inwards. Per unit $\beta$ the shift is $-3.9\times10^{-6}$, $-7.7\times10^{-5}$ and $-4.2\times10^{-4}$ at the $g=1$ threshold and $-9.4\times10^{-7}$, $-2.5\times10^{-5}$ and $-2.2\times10^{-4}$ at the $g=2$ threshold, for $m_X/\mu=1$, $0.2$ and $0.05$. The same shift changes $\kappa_g$ for every open harmonic. Its effect on the dephasing of section~\ref{sec:observables} is below $0.2\%$ of the linear term.

\subsection{Conservative orbital correction}
\label{app:conservative}
The static Yukawa interaction also induces a conservative
potential between the cloud and the companion,
\begin{equation}
U_X(R)=-\beta q\alpha^2 M_c X(x,\zeta).
\label{eq:C7}
\end{equation}
Relative to the Newtonian gravitational potential energy $U_N=-qM^2/R$, its magnitude is
\begin{equation}
\left|\frac{U_X}{U_N}\right|
=|\beta|\frac{M_c}{M}xX(x,\zeta).
\label{eq:C8}
\end{equation}
For the benchmark parameters, this correction changes the linear phase coefficient $\Phi_1$ by at most 1.2\%.
The effect of the static bound-state energy shift and
the associated threshold corrections has already been estimated in eqs.~\eqref{eq:C1}--\eqref{eq:C6}, and changes $\Phi_1$ by less than 0.2\%.
Both corrections are therefore subleading in the parameter range considered.

\section{Domain of validity}

\subsection{Companion perturbativity and Yukawa regime}
\label{app:perturbativity}

The benchmark uses a mass ratio $q=10^{-3}$, for which the isolated hydrogenic cloud provides a controlled starting point. Extending the analysis to $q\sim0.1$ would require accounting for cloud deformation, Roche-lobe effects and comparable-mass dynamics.

The static Yukawa shift of the $|211\rangle$ binding energy, quantified by $\epsilon_{\rm static}$ in appendix~\ref{app:static}, remains below $1\%$ at both termination thresholds considered in section~\ref{sec:bb}.  The ratio $|\eta_{200,211}|/|\Delta|$ is $8\times10^{-3}$ at the $211\to200$ threshold and $1.4\times10^{-4}$ at the continuum-ionization threshold. Near the companion, the distortion of the final states by $V_X$ is controlled by the Born parameter
$\mu|\beta|q\alpha/m_X$. For $m_X=\mu$, it is $0.021$ at the continuum threshold $|\beta|\simeq106$ and $0.33$ at the $211\to200$ threshold $|\beta|\simeq1.6\times10^3$.  The former is negligible, whereas the latter is not parametrically small. The $211\to200$ threshold should therefore be regarded as an estimate. First-order perturbation theory is otherwise adequate in the parameter range considered.

For an isolated attractive Yukawa potential around the companion, a local $s$-wave bound
state first appears when
\begin{equation}
\frac{2\mu\beta q\alpha}{m_X}\gtrsim1.680,
\qquad \beta>0.
\label{eq:local_yukawa_bound}
\end{equation}
This threshold refers to the isolated Yukawa potential and serves as a reference for the
two-center problem. Equivalently, $\mu|\beta|q\alpha/m_X\gtrsim0.84$ for the attractive sign. Beyond this threshold, perturbative ionization should be replaced by a two-center treatment that includes capture and multiple scattering.

\subsection{Retardation and real mediator emission}
\label{app:retardation}

The instantaneous form of the Yukawa potential in eq.~(3.4) neglects the frequency dependence of the massive propagator. For a harmonic $g$, the source satisfies
\begin{equation}
[\nabla^2-(m_X^2-g^2\Omega^2)]\varphi_g
=-(\mathrm{source})_g.
\label{eq:retarded_field}
\end{equation}
For $g\Omega<m_X$, the screening scale is therefore
$\sqrt{m_X^2-g^2\Omega^2}$, whereas for
$g\Omega>m_X$ the companion can emit an on-shell mediator.

At $x=6$, the latter condition requires $g\geq368,\ 74,\ 19$ for $m_X/\mu=1,\ 0.2,\ 0.05$, respectively, well above the harmonics carrying most of the power. Replacing $m_X$ by $\sqrt{m_X^2-g^2\Omega^2}$ harmonic by harmonic changes $P_X$ by $0.02\%,\ 0.23\%,\ 0.65\%$, and $P_{\rm int}$ by $0.006\%,\ 0.10\%,\ 0.32\%$, respectively, at $x=6$. Near the low-mass edge of figure~\ref{fig:decomp}, the correction to $P_X$ reaches $0.9\%$. Results in this region should therefore be regarded as estimates until the frequency-dependent retarded Green function is included.

\subsection{Additional limitations}
\label{app:limitations}
Equation~\eqref{eq:VX} treats the companion as a point-like Yukawa source. This approximation assumes that the companion's size is much smaller than the cloud size and the mediator range. Finite-size effects, screening, or spontaneous scalarization~\cite{DamourEF1993} could modify the effective source.

We assume that earlier Bohr and hyperfine resonances~\cite{Baumann2019,BBST2022prl} have not depleted or transferred the $|211\rangle$ cloud before the companion
reaches the orbital separations considered. Eccentricity and inclination, which can modify the ionization power~\cite{TSB2023}, are neglected.

The ionization power in eq.~\eqref{eq:power} already includes wave-mechanical dynamical friction~\cite{TSB2023,Traykova2021} and should not be supplemented by a separate
transport-cross-section estimate. Bound-state transitions, companion absorption, and on-shell mediator radiation are distinct channels not
included in this power.

Finally, our treatment uses non-relativistic continuum states of the 1/r gravitational potential and does not capture the full Kerr quasibound-state spectrum. Relativistic
studies have also identified limitations of non-relativistic estimates of environmental level shifts and mixing~\cite{Guo2026}. Our results should therefore be regarded as non-relativistic benchmarks rather than strong-field predictions.


\end{document}